\documentclass[sigconf, authorversion=true]{acmart}
\usepackage{todonotes}
\usepackage{float}
\usepackage{soul}
\setcopyright{none}

\usepackage{placeins}
\makeatletter
\def\@copyrightspace{\relax}
\makeatother
\renewcommand\footnotetextcopyrightpermission[1]{}

\acmConference[MIT Conference on Digital Experimentation]{}{November 1-2, 2019}{Cambridge, MA}

\begin{document}

\title[Online Surveys and Digital Demography in Kenya]{ Online Surveys and Digital Demography in the \\  Developing World: Facebook Users in Kenya} %

\author{Katherine Hoffmann Pham}
\affiliation{%
  \institution{Stern School of Business \\ New York University}
  \city{New York}
  \country{USA}
}
\author{Francesco Rampazzo}
\affiliation{%
 \institution{University of Southampton  \\
 and Max Planck Institute for Demographic Research}
  \city{Southampton}
  \country{United Kingdom \\~}
}
\author{Leah R. Rosenzweig}
\affiliation{%
 \institution{Institute for Advanced Study  \\
  and MIT GOV/LAB}
  \city{Toulouse}
  \country{France}
}


\maketitle

\section{Motivation}
Digital platforms have transformed the study of human behavior. Large online platforms such as Facebook, Twitter, and Wikipedia enable researchers to test behavioral theories on user communities ``in the wild'', whereas crowdsourcing platforms such as Amazon Mechanical Turk (MTurk) offer new opportunities to create synthetic online laboratory environments.	These platforms have broadened the subject pools available for academic research, allowing investigators to study more diverse populations than the conventional university undergraduate samples.

While the use of these online platforms to recruit subject pools has been well documented in the US and Europe, there is less evidence on whether these platforms can also be used to reach respondents in developing countries. As more of the world's citizens ``get online'', we expect that they will be easier to count, interview, and represent with digital platforms and survey research. Nevertheless, limited internet availability and computer literacy may restrict access to these platforms, raising the question of who, exactly, is represented. To investigate this question, we focus on the Facebook advertising platform and begin with a case study of Kenya. 

Facebook's advertising platform has gained traction among social science researchers as a result of two core capabilities. First, Facebook allows advertisers to produce ``audience estimates'' of the number of platform users with given demographic or behavioral characteristics. The potential of these estimates has been recognized by demographers, who have used this tool to study topics such as the digital gender divide and international migrant populations \cite{fatehkia_using_2018, zagheni_leveraging_2017}. %
  Second, Facebook allows advertisers to reach these tailored audiences with targeted ads. Researchers have used this capability to recruit respondent samples for opinion and attitude surveys in the US, India, and Brazil \cite{samuels_using_2013, boas_recruiting_2018, sances_ideology_2018}. To our knowledge, however, no research has systematically investigated the potential of Facebook advertising for representative research in an African country. %

We assess the value of these tools in Kenya using two primary strategies. First, we collect subnational Facebook audience estimates and compare these with official population data from the 2009 census. Second, we use Facebook's targeting tools to recruit a pilot sample of 957 %
respondents, whose demographics and survey responses we compare to the results of the 2016 Afrobarometer (Round 7) and the results of a canonical choice experiment (Tversky and Kahneman's 1981 Asian disease problem). %
This pilot data was collected in preparation for a full-scale survey that was launched in late September, immediately following the Round 8 Afrobarometer survey deployment and after the completion of Kenya's most recent census round in August 2019. 

\section{Related work}
\subsection{New data sources for the developing world}
While the use of Facebook to recruit respondents in the developing world is relatively new, researchers have explored how other big data sources might supplement costly or infrequent survey data collection in resource-poor settings. Early work focused on mobile phone data, using Call Data Records (CDR) to study phenomena such as the distribution of poverty, the spread of disease, or population mobility in response to natural disasters \cite{lu_predictability_2012, wesolowski_quantifying_2012, blumenstock_predicting_2015}. More recently, research has focused on remote sensing, including satellite and nighttime lights imagery, to study poverty and even political favoritism \cite{hodler_regional_2014,  jean_combining_2016}. Along these lines, we compare passively collected information from a novel data source (Facebook) to official statistics that are typically considered to be ``ground truth''. Rather than simply observing behavior remotely, we also have the ability to reach out to study subjects and survey them directly.

\subsection{The representativeness of online platforms}
This research also draws on work on the demographics of online platforms and how online samples compare to more established data sources. These studies have drawn attention to a number of important considerations in the use of online platforms to recruit survey respondents. For example, \citet{zhang_quota_2018} explore the potential of using targeted ads to generate representative surveys on attitudes towards climate change. They conduct stratified sampling of Facebook audiences in the US using age, education, gender, race, and region. While their results approximate their ``ground truth'' data sources, they note two key limitations of the Facebook data: first, they were unable to recruit any respondents in 157 of their 544 strata; and second, they observed within-strata imbalances %
on other characteristics (such as political orientation), which they attribute to the fact that Facebook strategically serves ads to those who are most likely to respond, exacerbating selection bias.

In the context of crowdsourcing,  \citet{difallah_demographics_2018} note a similar challenge with selection bias on the MTurk platform, since respondents may elect tasks that appeal to them. They also observe that the population on the platform is dynamic over time, with the entry and exit of workers; similar patterns may appear on Facebook, as users join or leave the platform.

A final interesting challenge has been raised by \citet{goel_online_2017} when conducting a comparison between MTurk and established survey sources. The authors observe that on a series of opinion questions, there are notable discrepancies \textit{between} survey sources, making it difficult to establish a ``ground truth'' to use as a benchmark. In such settings, online platforms may provide valuable additional insight into conflicting survey results. %

Despite these concerns, \citet{zhang_quota_2018} and \citet{goel_online_2017} find promising results: given the appropriate stratification and/or post-stratification weighting, data collected from online platforms is often able to approximate established, nationally representative data sources. A key question inspiring our research is whether the limitations described above are prohibitive in the context of developing countries like Kenya. %

\section{Context}
\begin{figure}
\centerline{\includegraphics[height=1.75in]{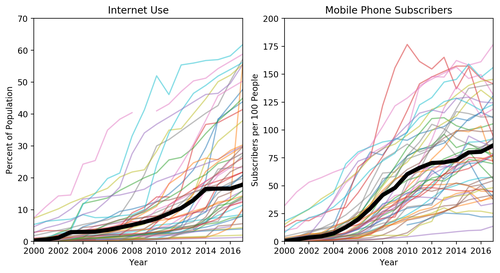}}
\caption{Annual growth in internet users and mobile phone subscribers in Africa; Kenya is shown in black. Source: \citet{international_telecommunications_union_statistics_2019}.}\label{fig:tech_use}
\end{figure}

We selected Kenya as a case study for two main reasons. First, Kenya represents an average example of mobile and internet use on the African continent (see Fig. \ref{fig:tech_use}). Kenya's telcom networks have seen a strong demand for mobile services: Safaricom's popular M-Pesa platform was a leading pioneer of mobile money, and Airtel participates in Facebook Free Basics to offer costless browsing on the social network. However, Kenya also has large populations without regular access to the internet or even electricity (e.g., 52\% of Afrobarometer respondents do not have a connection from the main electrical grid to their home), raising the question: how representative is the Facebook audience?

\begin{figure}
\centerline{\includegraphics[height=1.75in]{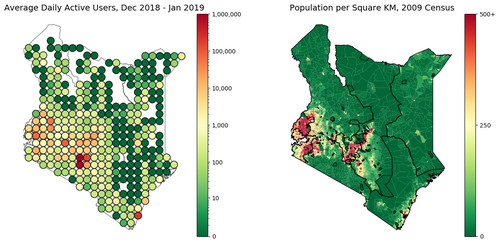}}
\caption{A comparison between Facebook's daily active user estimates for late 2018/early 2019, and population density estimates from the 2009 census. Missing circles indicate that Facebook estimated no users.} \label{fig:pop_density}
\end{figure}

Second, Kenya presents an excellent opportunity of timing. Specifically, two major survey efforts coincided: the decennial census from 24-31 August 2019, and the Round 8 Afrobarometer in September 2019. We planned our Facebook survey for September 2019, ensuring that any differences with the census and Afrobarometer would not be due to changes in the population or attitudes over time. While Facebook studies should not replace such rigorous, nationally representative surveys, these surveys are infrequently collected\footnote{Kenya's last census was in 2009, and Kenya's most recent Afrobarometer rounds were in 2011, 2014, and 2016.} and often delayed in their release. In contrast, Facebook data can be collected in near-real time. Thus, we examine the extent to which Facebook data collection can be used to complement existing sources with more frequent collection. %

\section{Data collection}
\subsection{Facebook audience estimates}
Audience estimate queries were submitted to the Facebook Marketing API using the PySocialWatcher Python package \cite{araujo_using_2017, araujo_social_2019}.
We collected %
estimates for Facebook users using two %
 key geographic targeting strategies. 
First, %
we divided the country into a \textit{grid of non-overlapping circles} with 20km radii, and collected audience estimates of Facebook users within each radius. This allowed us to examine geographic variation in the density of Facebook users across the country.
Second, %
 we generated estimates of the number of users within a 20km radius of each of \textit{Afrobarometer's survey clusters}. This enabled us to study the number of Facebook users in areas targeted by the Afrobarometer, in order to understand whether Facebook samples could feasibly be recruited there.

\subsection{Facebook surveys}
To supplement these audience estimates and understand how they translate into researchers' ability to engage users, we also conducted a pilot survey on the Facebook advertising platform. Between 6 and 9 June, we advertised the opportunity to participate in a 15-minute survey in exchange for a 50 Ksh (about 50 cents) mobile airtime credit. During this period, we were able to recruit 1,190 Kenyans on Facebook to take the survey, with a total cost of \$500 in advertising fees and \$630 in respondent incentives and administrative costs. After removing duplicate respondents, we have a sample of 957 unique respondents who spent a reasonable amount of time on the survey ($\geq$ 5 minutes). 

We explored two different ad targeting strategies. First, we targeted advertisements to users residing in each of Kenya's eight \textit{provinces}. This approach was designed to force geographic variation in respondent locations, while minimizing the number of separate campaigns. %
Second, we targeted advertisements to users residing within a 40km radius of select \textit{Afrobarometer clusters}. For each province, one Afrobarometer cluster was selected as a median-case example cluster based on the audience estimates provided by the Facebook API. Furthermore, we selected two worst-case clusters for which Facebook estimated less than 1,000 monthly active users (and zero daily active users). This approach was designed to explore the cost and difficulty of recruiting respondents near these clusters.

\begin{figure}
\centerline{\includegraphics[width=3.3in]{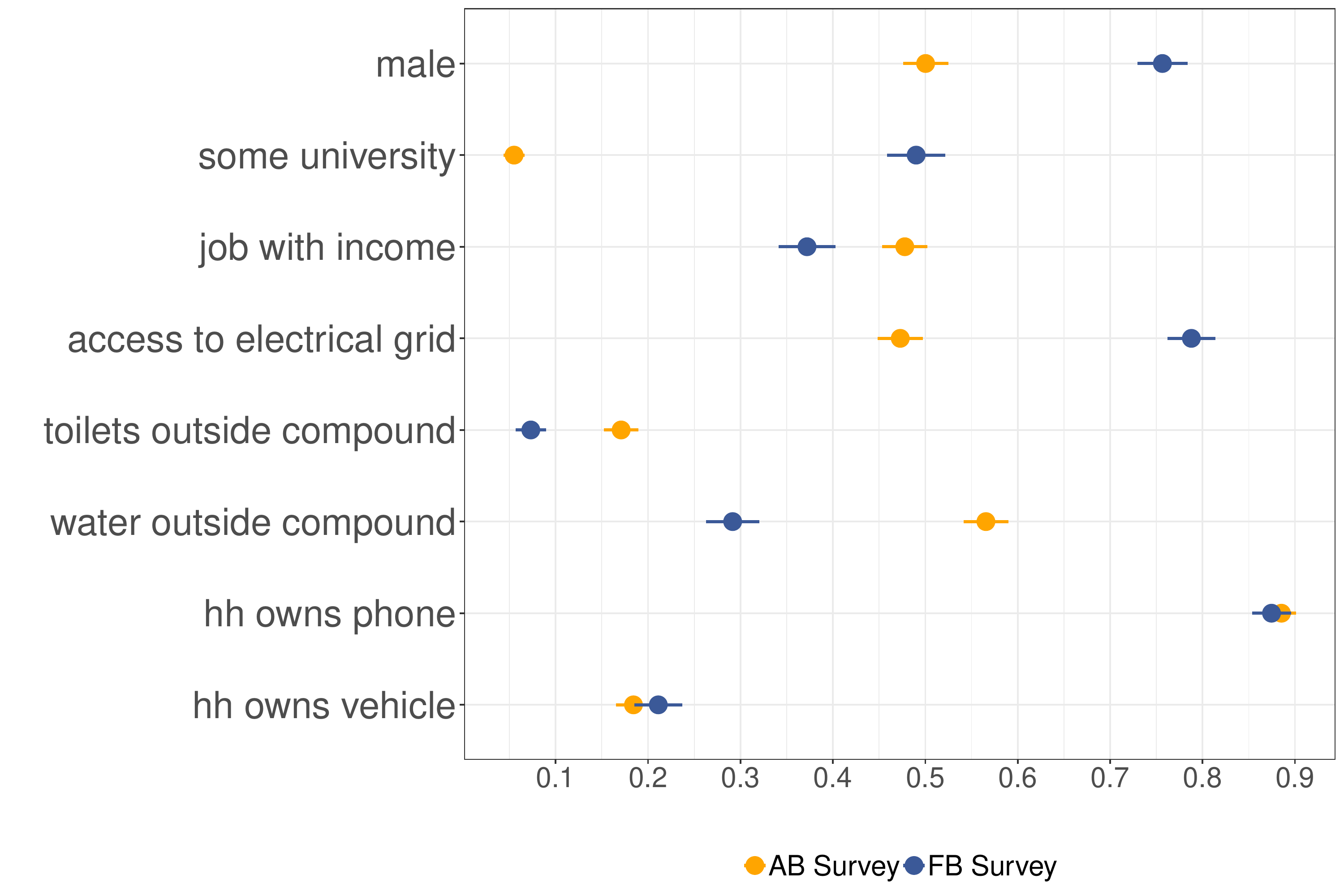}}
    \caption{A comparison of demographics between Afrobarometer and our Facebook sample}
    \label{fig:demsplot}
\end{figure}

\section{Preliminary Results}
\subsection{Facebook audience estimates}

As shown in Fig. \ref{fig:pop_density}, Facebook has an active user base across most of the country. Facebook audience sizes generally reflect underlying population density: there are many users in the Nairobi and Mombasa areas, as well as in the population-dense Western parts of the country. Unsurprisingly, Facebook is underrepresented along Kenya' northern borders, which consist of relatively remote and dry regions shaped in part by the military conflict with Somalia. %

\subsection{Facebook surveys}

\subsubsection{Logistical aspects of recruiting Facebook respondents}
Our advertisements reached 159,743 Facebook users in a four-day period. The performance of our advertisements varied by province. Unsurprisingly, users in Nairobi were responsive, with a click-through rate of 5\%, a completion rate of 40\%, and an advertising cost of \$0.05 per survey. Our worst performance was in North Eastern province, with a 1\% click-through rate, a 7\% survey completion rate, and an advertising cost of \$2.65 per survey. 
Targeting at the cluster level increased advertising costs. Across the test clusters with median audience sizes, these costs ranged from \$0.36 per survey in Rift Valley to \$9.71 in the North East. In one cluster -- the worst-case Rift Valley cluster -- we were unable to recruit any respondents. %

\subsubsection{Geography of Facebook respondents}
Although we used a limited targeting strategy for the pilot, respondents reported coming from all 47 counties of Kenya, and had 13 different tribal/ethnic affiliations. %
Generally speaking, respondents' self-reported locations did \textit{not} closely match a given ad's target location. It is possible that respondents have ties to multiple locations; we have added additional survey questions to the full-scale survey to address this possibility. However, others have documented similar issues with Facebook's geographic ad targeting in the US context \cite{sances_missing_2019}.

\subsubsection{Demographics of the Facebook respondents}
Relative to the Afrobarometer sample, our respondents were more likely to be male, university-educated, and connected to the electrical grid. They were less likely to be employed, or to leave their compound to access water or sanitation. We identified comparable rates of phone and vehicle ownership across the samples (see Fig. \ref{fig:demsplot}).

\subsubsection{Experimental results} %
\begin{figure}
\centerline{\includegraphics[width=3.3in]{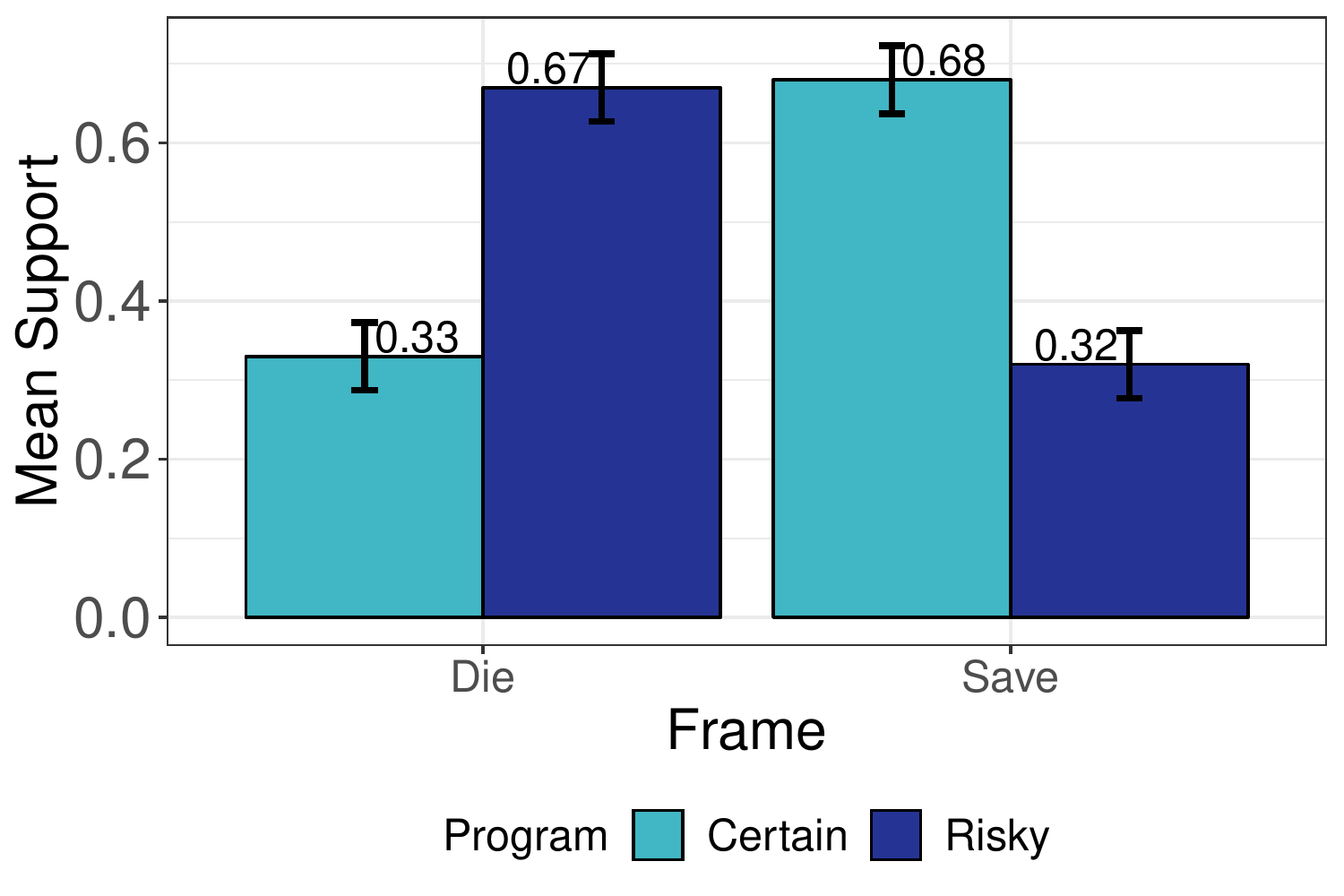}}
    \caption{Replication of the Tversky \& Kahneman 1981 Asian Disease Experiment}
    \label{fig:asiandisease}
\end{figure}

We replicated a canonical behavioral experiment --- Tversky and Kahneman's 1981 Asian disease problem --- that has been conducted in a wide range of contexts over time \cite{tversky_framing_1981}. Interestingly, our results are similar to %
the original sample and other %
Western convenience samples \cite{berinsky_evaluating_2012}. When the problem is framed in terms of the number of people who will die from the disease, a majority (67\%) of respondents prefer the risky option. When the problem is framed in terms of the number of lives %
saved, a similar majority (68\%) prefer the certain policy (see Fig. \ref{fig:asiandisease}). 

\section{Conclusion and Next steps}
We conduct one of the first systematic studies of Facebook as a research tool in the developing world. We assess the potential of Facebook for producing population estimates and recruiting survey respondents. We present preliminary results from a pilot survey of 957 respondents, which suggest that Facebook can be used to recruit respondents from diverse areas of Kenya at reasonable costs. Although the sample exhibits predictable biases in terms of gender, access to infrastructure, and education levels, our sample is less privileged than expected: over one fifth of respondents said that their home did not have a connection to the electrical grid, and over one quarter did not have piped water inside their home.

Our full-scale survey tests %
whether a more sophisticated sampling strategy, involving stratification on gender and 25 separate regional clusters, will address some of the biases in our sample. We also plan %
to explore whether post-stratification can further reduce remaining imbalances in demographics, before using the weighted sample to compare attitudes and behaviors with those from nationally representative surveys. %
Finally, we are also considering different strategies for extending our sample. These could include standard approaches such as respondent-driven sampling, or developing a custom objective function for the ad targeting algorithm, similar to the approach of \citet{ipeirotis_quizz:_2014}.

Digital inclusion and social media offer exciting new opportunities for experimental and survey research in the developing world. We build a foundation for such research by documenting the type of respondent sample that can be recruited via an online platform. We expect that this will form part of a larger line of research describing the strengths, limitations, and biases of these new tools. In general, we are optimistic about the use of these platforms for social science research. %
We anticipate that our findings will be of broad interest to researchers working in developing countries where an increasing number of their populations of interest are accessing social media.

\section{Acknowledgements}
We thank the Summer Institute for Computational Social Science (SICSS), the Russell Sage Foundation, and the Alfred P. Sloan Foundation for financial support. We thank Chris Bail, Matt Salganik, Anne Helby Petersen, Julien Migozzi, and Tina Law for their assistance in shaping this project. We thank Ingmar Weber, Alessandro Sorichetta, Dennis Feehan, and researchers at the Busara Center for Behavioral Economics and Afrobarometer for helpful comments. We thank Nelson Ngige, Eunice Williams, and Kibuchi Eliud, Warsama Abdifitah, and Ahmed Hared for translation assistance.

\bibliographystyle{ACM-Reference-Format}
\bibliography{literature}
\end{document}